\documentstyle[aps,preprint,epsfig]{revtex}
\tightenlines
\def\beq{\begin{equation}}
\def\eeq{\end{equation}}
\def\beqa{\begin{eqnarray}}
\def\eeqa{\end{eqnarray}}
\def\MeV{\nobreak\,\mbox{MeV}}

\def\fm{\nobreak\,\mbox{fm}}

\def\ii{\'{\i}}
\def\nn{\nonumber}

\begin{document}
\draft
\title{Droplet formation in cold asymmetric nuclear matter in the
quark-meson-coupling model \\}
\author{G. Krein$^{1,2}$
\thanks{e-mail: gkrein@ift.unesp.br}
D.P. Menezes$^3$
\thanks{e-mail: fsc1dpm@fsc.ufsc.br}  
M. Nielsen$^4$
\thanks{e-mail: mnielsen@if.usp.br} 
and C. Provid\^encia$^5$
\thanks{e-mail: cp@teor.fis.uc.pt}
\\}
\address{1. Department of Physics, University of Washington, Box 351560, 
Seattle - WA 98195-1560, USA }
\address{2. Instituto de F\ii sica Te\'orica, Universidade Estadual Paulista \\
R. Pamplona, 145 - 01405-900 S\~ao Paulo, SP, Brazil}
\address{3. Departamento de F\ii sica - CFM,  Universidade Federal de Santa 
Catarina \\
Caixa Postal 476, 88.040-900 Florian\'opolis, SC, Brazil}
\address{4. Instituto de F\ii sica, Universidade de S\~ao Paulo \\
Caixa Postal 66318, 05315-970  S\~ao Paulo, SP, Brazil}
\address{5. Centro de F\ii sica Te\'orica - Departamento de F\ii sica,
Universidade de Coimbra \\
3000 - Coimbra - Portugal }

\maketitle

\begin{abstract}
The quark-meson-coupling model is used to study droplet formation from the 
liquid-gas phase transition in cold asymmetric nuclear matter.  The critical 
density and proton fraction for the phase transition are determined in the
mean field approximation. Droplet properties are calculated in the 
Thomas-Fermi approximation. The electromagnetic field is explicitly included 
and its effects on droplet properties are studied. The results are compared
with the ones obtained with the NL1 parametrization of 
the non-linear Walecka model. 
\end{abstract}

\vspace{1.0cm}

\newpage
\section {Introduction}

One of the most important problems in contemporary nuclear physics and 
astrophysics is the determination of the properties of nuclear matter as 
functions of density, temperature and the neutron-proton composition. 
In fact, neutron-star matter at densities between $0.03~\fm^{-3}$ and nuclear 
matter saturation density consists of neutron-rich nuclei immersed in a gas 
of neutrons. In~particular, understanding the transition crust-core in neutron
stars is essential for explaining a number of properties of these 
stars~\cite{pr}. To achieve this goal, one must study not only the ground and 
excited states of normal nuclei, but also nuclear states of high excitation 
and far from stability. 

Recently, two of us studied droplet formation from the liquid-gas phase 
transition in cold~\cite{mp1} and hot~\cite{mp2} asymmetric nuclear matter 
in the context of the non-linear Walecka model (NLWM)~\cite{bb}. 
In the present paper
we employ the  quark-meson-coupling (QMC) model, originally proposed by 
Guichon~\cite{guichon}. In the QMC model, the mean scalar $\sigma$ and vector 
$\omega$ meson fields couple directly to the confined quarks inside the 
nucleon bags, instead to point-like nucleons as in Walecka-type 
models~\cite{SW}. The saturation of nuclear matter in this model is due to 
the density dependence of the effective $\sigma N N$ coupling, generated by 
the coupling of the mean $\sigma$ field to the quarks. As we shall show in 
this paper, this density dependence of the coupling has important consequences
for the phase-transition and for the droplet radius and surface energy 
density.
While the QMC model shares many similarities with Walecka-type 
models~\cite{SW}, it however offers new opportunities for studying nuclear 
matter properties. Perhaps one of the most exciting ones is the possibility 
of using the same model to study nuclear phenomena in a large range of 
densities. With the QMC model, we have the opportunity to investigate the 
density regime where the quarks 
remain confined in the nucleon bags, but the stucture of the nucleons 
nevertheless changes, as became evident from the EMC effect~\cite{EMC}. We 
also expect to use the QMC model at much higher densities, where the nucleon 
bags start loosing their identity and the deconfinement transition starts
taking place~\cite{deconf,highden}. It is therefore important to explore the 
performance of the model in all such density regimes. Another good reason to 
use the QMC model in the confined regime is that it might be of help to fix 
the low-energy constants of relativistic Lagrangians, where nucleon 
substructure is incorporated through a derivative expansion~\cite{EFT}.   

Since the original version of the model, there have  
been several ameliorations and extensions. These include the treatment of the 
nucleon center-of-mass 
motion~\cite{Fmot,st1}, treatment of finite nuclei~\cite{fnuc1,fnuc2,fnuc3}, 
change of the bag constant in medium~\cite{Bstar1}, inclusion of finite 
temperature effects~\cite{finT} and the treatment of Fock~\cite{Fock} 
and quark-exchange~\cite{Qexch} terms. For a list of applications of 
the model to a 
variety of nuclear phenomena and earlier studies within the QMC model,
see Ref.~\cite{AWT}. 

Within the framework of relativistic models, the liquid-gas phase transition 
in nuclear matter has been investigated at zero and finite temperatures for 
symmetric and asymmetric semi-infinite systems~\cite{rs,md,epsw,cev} and for 
finite systems \cite{mp1,mp2}. The present study aims to study the liquid-gas 
phase transition and droplet formation in a vapor system at zero temperature 
in the context of the QMC model and explore  differences from the
results of Refs.~\cite{mp1,mp2}. We include the Coulomb interaction and work 
in the Thomas-Fermi approximation. We determine the conditions for phase 
coexistence in a binary system by building the binodal section of the QMC 
model at zero temperature. As shown in Refs.~\cite{mp1,mp2}, the optimal 
nuclear size of a droplet in a neutron gas is determined by a delicate balance
between nuclear Coulomb and surface energies. The surface energy favors nuclei
with a large number of nucleons $A$, while the nuclear Coulomb self-energy 
favors small nuclei. However, because of the density dependence of the 
effective $\sigma NN$ coupling in the QMC model, the critical pressure and
proton fraction of the phase transition and the droplet properties turn out 
to be significantly different from the 
NL1 parametrization of the  NLWM\cite{bb}. We have chosen to compare 
the QMC model results with the ones obtained with the NL1 parametrization 
because this parametrization has proven to give an excellent reproduction 
of the ground-state properties of the nuclei in general~\cite{bb,pgr}. 
The values calculated for the QMC model are also parameter dependent, but 
other possible parametrizations produce the same qualitative features as
the one chosen in this work. We also see that, as expected, the presence of
the outside neutron gas reduces the surface tension, since as the density of 
the system increases, the inside and outside matter become more and more 
alike. 

The paper is organized as follows: in Sect.~II we  briefly summarize
the QMC model for finite asymmetric systems including the couplings of 
the $\rho$ meson and the photon to the quarks. In Sect.~III we discuss
the Thomas-Fermi approximation and in Sect.~IV we apply the model to
asymmetric  nuclear matter. Finally in Sects.~V and VI we give our
numerical results and conclusions.

\section{The QMC model for finite asymmetric systems}

In a nucleus, the motion of the nucleon is relatively slow and the quarks 
are highly relativistic. The internal structure of the nucleon has therefore 
enough time to adjust to the external local meson fields, the motion of the 
nucleon can be treated as a point-like Dirac particle~\cite{fnuc1,fnuc2,fnuc3}
with an effective mass $M^*_N$. The effective mass depends on the position 
only through the mean scalar field $\sigma$. To describe a finite system with 
different numbers of protons and neutrons, it is necessary to consider also the
contribution of the $\rho$ meson. Besides, any realistic treatment of 
nuclear structure also requires that one introduces the Coulomb
force. Therefore, a possible Lagrangian density for a static system is
\beqa
{\cal L}_{QMC}&=& \bar{\psi} [i \gamma \cdot \partial 
- M_N^*(\sigma({\mathbf{r}}))   
- g_\omega \omega({\mathbf{r}}) \gamma_0 
- g_\rho \frac{\tau^N_3}{2} b({\mathbf{r}}) \gamma_0 
- \frac{e}{2} (1+\tau^N_3) A({\mathbf{r}}) \gamma_0 ] \psi \nn \\
&-& \frac{1}{2}[ (\nabla \sigma({\mathbf{r}}))^2 + 
m_{\sigma}^2 \sigma({\mathbf{r}})^2 ] 
+ \frac{1}{2}[ (\nabla \omega({\mathbf{r}}))^2 + m_{\omega}^2 
\omega({\mathbf{r}})^2 ] \nn \\
&+& \frac{1}{2}[ (\nabla b({\mathbf{r}}))^2 + m_{\rho}^2 b({\mathbf{r}})^2 ] 
+ \frac{1}{2} (\nabla A({\mathbf{r}}))^2 \; ,
\label{lag}
\eeqa
where $\psi({\mathbf{r}})$, $\sigma({\mathbf{r}})$, $\omega({\mathbf{r}})$,  
$b({\mathbf r})$ and $A({\mathbf r})$ are respectively the nucleon,  meson 
$\sigma$ and the mean values of the time component of $\omega$, $\rho$ and 
Coulomb fields in the nucleon rest frame;  $\tau^N_3/2$ is the third 
component of the nucleon isospin operator; $m_\sigma,\;m_\omega$ and $m_\rho$ 
are respectively the masses of the $\sigma,\;\omega$ and $\rho$ fields; 
$g_\omega$ and $g_\rho$ are the $\omega-N$ and $\rho-N$ coupling constants, 
which are related to the corresponding quark coupling constants as
$g_\omega=3g_\omega^q$,  $g_\rho=g_\rho^q$. Finally, $e$ is the electric 
charge.

The effective nucleon mass $M_N^*$ is calculated in the MIT bag model. 
Parametrizing the sum of the center-of-mass and gluonic fluctuation 
corrections in the familiar form $-z/R$, where $R^*$ is the in-medium bag 
radius, $M_N^*$ takes the form 
\beq
M_N^*={3\Omega_q - z\over R^*}+{4\over3}\pi B R^{*3} ,
\label{mn}
\eeq
where $\Omega_q=\sqrt{x^{*2}+(R^* m^*_q)^2}$ is the kinetic energy of the 
quarks, $m_q^*({\mathbf{r}})=m_q-g_\sigma^q\sigma(\mathbf{r})$ is the effective
quark mass, $m_q$ is the bare quark mass,  $g_\sigma^q$ 
is the quark-$\sigma$ coupling constant, and $x^*$ is the in-medium bag 
eigenvalue determined from the boundary condition
\beq
j_0(x^*)=\sqrt{\Omega_q-R^* m^*_q\over\Omega_q+R^* m^*_q} \; j_1(x^*)\,.
\eeq
The bag radius is obtained by minimizing $M_N^*$ with respect to $R^*$
\beq
{\partial M_N^*\over\partial R^*}=0\,,
\label{sta}
\eeq
and the bag constant $B$ and the parameter $z$ are fixed to reproduce the 
free-space nucleon mass ($M_N=939~\MeV$). In this work we use $m_q=5~\MeV$ 
and fix the free bag radius at $R_B=0.8~\fm$. The results for $B$ and $z$ 
are: $B^{1/4}=170.0~\MeV$ and $z=3.295$. Results for $B$ and $z$ for other 
values of the bare quark mass and bag radius can be found in 
Ref.~\cite{fnuc3}.

The variation of the Lagrangian, Eq.~(\ref{lag}), results in the
following equations for a spherically symmetric system
\beqa
\frac{d^2}{dr^2} \sigma(r) + \frac{2}{r} \frac{d}{dr} \sigma(r) 
    - m_\sigma^2 \sigma(r) &=& - g_\sigma C(\sigma(r)) \rho_s(r) \;,
\label{sig} \\
\frac{d^2}{dr^2} \omega(r) + \frac{2}{r} \frac{d}{dr} \omega(r) 
    - m_\omega^2 \omega(r) &=& - g_\omega \rho_B(r) \;,
\label{omg} \\
\frac{d^2}{dr^2} b(r) + \frac{2}{r} \frac{d}{dr} b(r) 
    - m_\rho^2 b(r) &=& - \frac{g_\rho}{2} \rho_3(r) \;,
\label{rho} \\
\frac{d^2}{dr^2} A(r) + \frac{2}{r} \frac{d}{dr} A(r) 
    &=& - e \rho_p(r) \;,
\label{cou} 
\eeqa
where
\beq
g_\sigma C(\sigma) =-{\partial M_N^*\over\partial\sigma} =
3 g_\sigma^qS(\sigma)\;,
\label{gC}
\eeq
with
\beq
S(\sigma) = \frac{\Omega_q/2 + m_q^{*}R_B(\Omega_q-1)}
{\Omega_q(\Omega_q-1) + m_q^{*}R_B/2}\;. 
\eeq
Note that the $\sigma-N$ coupling constant $g_\sigma$ is related to the
quark-$\sigma$ coupling $g_\sigma^q$ through the relation
\beq
g_\sigma=3 \, g_\sigma^q \, S(0)\,,
\label{ggq}
\eeq
and therefore
\beq
C(\sigma)={S(\sigma) \over S(0)}\,.
\eeq
Also, in Eqs.~(\ref{sig}) to (\ref{cou}), $\rho_s$ is the nucleon scalar 
density in the system with $A$ nucleons 
\beq
\rho_s(r)=\langle A|\left[\bar{\psi}_p(r) \psi_p(r) + 
\bar{\psi}_n(r) \psi_n(r)\right]|A\rangle \,, 
\label{scadens}
\eeq
$\psi_p$ and $\psi_n$  are the proton and neutron spinors, respectively,
and $\rho_B$ and $\rho_3$ are given in terms of the proton and neutron vector 
densities $\rho_p$ and  $\rho_n$
\beq 
\rho_B = \rho_p + \rho_n \,,\hspace{1.5cm}
\rho_3 = \rho_p - \rho_n , 
\label{vecdens}
\eeq
with
\beq
\rho_p = \langle A|{\psi}^{\dag}_p(r)\psi_p(r)|A\rangle, \hspace{1.0cm}
\rho_n = \langle A|{\psi}^{\dag}_n(r)\psi_n(r)|A\rangle \,.
\label{rhopn}
\eeq

\section {The Thomas-Fermi approximation}

Since we want to compare results with Refs.~\cite{mp1,mp2}, we employ the 
semi-classical Thomas-Fermi approximation, instead of solving the Dirac 
equation for the nucleons. This amounts to assuming that the mesonic fields 
vary slowly enough so that the baryons can be treated as moving in locally 
constant fields at each point of space.  The basic quantity in the Thomas-Fermi
approach is the phase-space distribution function for protons and neutrons 
\beq
f_i(r,k)=\theta\left(k_{F_i}(r)-k\right),\;\;\;\;i=p, n
\label{fi}
\eeq
where $k_{F_i}(r)$ is the local Fermi wave number for protons and
neutrons. The total energy of the system is given by
\beqa
E &=& 2\sum_{i=p,n}\int\,{d^3k \, d^3r\over(2\pi)^3} f_i(r,k) h_i(r,k) +
\frac{1}{2} \int d^3r
\left [ (\nabla \sigma)^2 +m_\sigma^2\sigma^2\right.\nn\\
&-&\left.(\nabla \omega)^2 - m_\omega^2 \omega^2 - (\nabla b)^2 - 
m_\rho^2 b^2 - (\nabla A)^2\right]\;,
\label{etot}
\eeqa
where $ h_i(r,k)=\sqrt{k^2+{M_N^*}^2(\sigma(r))} + \nu_i(r)$, with
\beqa
\nu_p(r)&=&g_\omega\omega(r)+{g_\rho\over2}b(r) + eA(r)~,\\
\nu_n(r)&=&g_\omega\omega(r)-{g_\rho\over2}b(r) \;.
\eeqa

The thermodynamic potential is defined as
\beq
\Omega=E-\sum_{i=p,n} \mu_i N_i\;,
\eeq
where $\mu_i$ is the chemical potential for particles of type $i$ and
$N_p$ and $N_n$ are, respectively, the numbers of protons and neutrons
\beq
N_i=\int d^3r \, \rho_i(r),\;\;\;\;i=p, n\;,
\eeq
with the proton and neutron vector densities, $\rho_p$ and $\rho_n$ of
Eq.~(\ref{rhopn}), given explicitly by
\beq
\rho_i(r)=2\int\,{d^3k\over(2\pi)^3} f_i(r,k)={1\over3\pi^2} k_{F_i}^3(r)\,.
\label{denp}
\eeq

Minimizing the thermodynamic potential $\Omega$ with respect to the
local Fermi momentum, the following expressions for the proton and neutron
chemical  potentials are obtained
\beqa
\mu_p&=&\sqrt{k_{F_p}^2+{M_N^*}^2}+ g_\omega \omega  + {g_\rho\over2}  b + 
e A\;,
\label{mup}\\
\mu_n&=&\sqrt{k_{F_n}^2+{M_N^*}^2}+ g_\omega \omega - {g_\rho\over2}  b\;,
\label{mun}
\eeqa
which can be used to find $k_{F_i}(M_N^*,\omega,b,A,\mu_i;r)$.

The fields that minimize $\Omega$ satisfy Eqs.~(\ref{sig}) to
(\ref{cou}), where the scalar density of Eq.~ (\ref{scadens}) is given explicitly 
by
\beq
\rho_s(r)=2\sum_{i=p,n}\int\,{d^3k\over(2\pi)^3}{M_N^*(\sigma(r))
\over\sqrt{k^2 + {M_N^*}^2(\sigma(r))}} f_i(r,k)\,.
\label{densc}
\eeq

\section{Liquid-gas phase transition in asymmetric nuclear matter}

Phase transitions in binary systems are more complex than in one-component 
systems because two kinds of instabilities can occur. We have therefore
two stability conditions. We have the condition for mechanical stability, 
which requires
\beq
\left({\partial P\over\partial\rho_B}\right)_{Y_p}\ge 0\;,
\eeq
where $P$ is the pressure and $Y_p=\rho_p/\rho_B$ is the proton fraction. 
We have also the condition for diffusive stability, which implies the 
inequalities
\beq
\left( \frac{\partial \mu_p}{\partial Y_p} \right)_{P} \ge 0
~~~{\rm and} ~~~
\left( \frac{\partial \mu_n}{\partial Y_p} \right)_{ P} \le 0\;.
\label{ds}
\eeq
These reflect the fact that in a stable system, energy is required to increase
the proton concentration while the pressure is kept constant.

In the mean field approximation for infinite nuclear matter, 
the meson fields are replaced by their expectation values and 
Eqs.~(\ref{sig}) to (\ref{cou}) become (omitting the electromagnetic field)
\beqa
\overline{\sigma}&=&\frac{g_\sigma }{m_\sigma^2}C(\overline{\sigma})\rho_s\;,
\label{sigf}\\
\overline{\omega}&=&\frac{g_\omega }{m_\omega^2} \rho_B\;, 
\label{omgf}\\
\overline{b}&=&{g_\rho\over2m_\rho^2}\rho_3\;,
\eeqa
where the sources of the fields are constants and can be related to
the nucleon Fermi momentum $k_{F_i}$ through equations (\ref{densc})
and (\ref{denp}), with the distribution functions given by (\ref{fi})
and the fields replaced by their expectation values.

%

Under this approximation, the energy density and pressure 
are given by

\beq 
{\cal E}=2 \sum_{i=p,n}\int {d^3k\over(2\pi)^3} \sqrt{k^2+{M_N^*}^2
(\overline{\sigma})}\; \theta (k_{F_i} - k)  
+{g_\omega^2\rho_B^2\over2m_\omega^2}+ 
{g_\rho^2\rho_3^2\over8m_{\rho}^2} 
+\frac{m_\sigma^2}{2}\overline{\sigma}^2\;,
\eeq
\beq
{ P}=\frac{2}{3} \sum_{i=p,n}\int {d^3k\over(2\pi)^3} {k^2\,\theta
  (k_{F_i} - k) \over\sqrt{k^2+{M_N^*}^2(\overline{\sigma})}}
+{g_\omega^2\rho_B^2\over2m_\omega^2}+ 
{g_\rho^2\rho_3^2\over8m_{\rho}^2} 
-\frac{m_\sigma^2}{2}\overline{\sigma}^2\;.
\eeq

There are six parameters to be determined: $g_\sigma$, $g_\omega$, 
$g_\rho$, $m_\sigma$, $m_\omega$ and $m_\rho$.  We take the 
experimental values of $m_\omega$ = 783 MeV and $m_\rho$ = 770 MeV,  and
the other parameters are taken from Ref.~\cite{fnuc3}: $g_\sigma^2/4\pi=3.12$,
$g_\omega^2/4\pi=5.31$, $g_\rho^2/4\pi=6.93$ and $m_\sigma=418$ MeV.

The two-phase liquid-gas coexistence is governed by the Gibbs condition
\beqa
\mu_i(\rho_p,\rho_n,M_N^*)&=&\mu_i(\rho_p^{\prime},\rho_n^{\prime},
{M_N^*}^{\prime}),~~i=p,n
\label{gi1}\\
{ P}(\rho_p,\rho_n,M_N^*)&=&
{ P}(\rho_p',\rho_n',{M_N^*}^{\prime})\;.
\label{gi2}
\eeqa
We have made use of the geometrical construction~\cite{barranco,ms} in order 
to obtain the chemical potentials in the two coexisting phases for each 
pressure of interest. In Fig.~1 we plot $\mu_p$ and $\mu_n$ as a function of 
the proton fraction for $P=0.1~\MeV/\fm^3$ (solid line). For comparison, we 
also show in this figure the results obtained with the linear Walecka 
model~\cite{sw79}
(dotted line) and with the NL1 non-linear  Walecka model~\cite{bb}, discussed 
in Refs.~\cite{mp1,mp2} (dashed line).

In order to obtain the binodal section, which gives the pairs of points under 
the same pressure for different proton fractions obtained using the geometrical
construction, we have used the condition of the diffusive stability, 
Eq.(\ref{ds}), and solved simultaneously Eqs.~(\ref{gi1}) and (\ref{gi2}), 
together with 
\beqa
\overline{\sigma}&=&\frac{g_\sigma}{m_\sigma^2}C(\overline{\sigma})\rho_s
(\rho_p,\rho_n,M_N^*) \,,
\\
\overline{\sigma}^{\prime}&=&\frac{g_\sigma }{m_\sigma^2}
C(\overline{\sigma}^{\prime})\rho_s(\rho_p',\rho_n',{M_N^*}^{\prime}) \,.
\eeqa

The binodal section  is shown in Fig.~2 for the QMC model (solid line). It is 
divided into two branches by the critical point (CP). One branch describes the
system in a high-density (liquid) phase, while the other branch describes the 
low-density (gas) phase. The liquid (gas) phase appears at the right (left) of CP.

For finite temperatures, $T\neq0$, the gas phase is also characterized by 
$Y_p\neq0$ for all values of the pressure. However, for $T=0$ and for 
 $P \lesssim 0.43$ MeV fm$^{-3}$, the gas phase always has $Y_p=0$. For the 
sake of completeness, we also show in Fig.~2 the binodal  sections for the 
NL1 discussed in refs.~\cite{mp1,mp2} (dashed line). We see that the QMC has 
a CP with higher values of $P$ and $Y_p$, but that in the liquid phase it is 
mainly characterized, for a given pressure $P$, by smaller values of $Y_p$ 
than the NL1.

\section{Numerical Results for Finite Systems}

The solution for the infinite system gives us the initial and
boundary conditions for the program which integrates the set of
coupled non-linear differential equations (\ref{sig}) to (\ref{cou})
in the Thomas-Fermi approximation. In this work the numerical
calculation was carried out with the iteration procedure described in
Ref.~\cite{mp1}, which uses also as input the size of the mesh,
$R_{mesh}$. The size of the mesh determines the size of the droplet
and, consequently, the chemical potentials and the number of particles 
in the droplet. Hence, the number of particles and the proton fraction
within the droplet are output of the program and not fixed from the start.
The same is true for the liquid and gas proton fractions. In order to
obtain a certain number of particles inside the droplet, we vary
$R_{mesh}$ for a given initial condition until the desired number is
obtained as output. Therefore, by fixing the number of the particles in the 
droplet we can study the behavior of some of its properties, such as the 
surface tension, neutron thickness, proton radius, etc., as a function of the 
central (r=0) proton fraction. For more details about this numerical
procedure we refer the reader to Refs.~\cite{mp1,mp2}.

The droplet surface energy per unit area in the small thickness
approximation is given by~\cite{mp1,mp2,np}
\beq
{\cal E}_{surf}=\int_0^\infty d r \left[ \left(\frac{d \sigma}{d r}\right)^2- 
\left(\frac{d\omega}{d r}\right)^2 -
\left(\frac{d b}{d r}\right)^2 \right].
\label{esup}
\eeq
In Ref.~\cite{mp1} the results obtained from this expression for the density 
surface energy were parametrized and compared with the liquid drop model 
results. The conclusion was that, albeit approximate, Eq.~(\ref{esup}) 
provides a very good estimate of the surface energy density.

The proton and neutron radii in spherical geometry, $R_i~(i=p,n)$, are
defined as 
\begin{equation}
\int_0^{R'} \rho_p(r) r^2 d r = \frac{1}{3}
\left[\rho_{p,g}R_p^3 + \rho_{p,l}({R'}^3-R_p^3) \right],
\label{raiosp}
\end{equation}
and 
\begin{equation}
\int_0^{R'} \rho_n(r) r^2 d r = \frac{1}{3}
\left[ \rho_{n,g} R_n^3+ \rho_{n,l}({R'}^3-R_n^3) \right],
\label{raiosn}
\end{equation}
where $\rho_{i,l}$ and $\rho_{i,g}$  refer to the liquid and gas density
respectively; $R'$ is the value of $r$ such that $|f(r)-f_g|< 10^{-8}$,
with $f$ being either a meson field or the baryonic density at $r$, and
$f_g$ the corresponding gas value. This means that $R'$ is the value
of $r$ for which the fields and density reach their asymptotic gas values.
 
Another important quantity is the thickness of the  region at the surface with
extra neutrons known as {\it neutron skin}. The {\it neutron skin thickness} 
is defined as~\cite{cev} 
\begin{equation}
\Theta=R_n-R_p.
\end{equation}

In order to understand the differences between the QMC and NL1
in the Thomas-Fermi approximation, we have calculated the proton density
of a droplet with total particle number $A=40$ and proton fraction
$N_p/A=0.5$. The results are shown in Fig.~3 and Table~I. The solid and dashed
lines in Fig.~3 show the results for QMC and NL1, respectively. It is clear 
from this figure that the QMC result shows higher central density and smaller 
proton radius. This feature is present in all solutions we have obtained. 
In average the central density is $\sim 0.01 -0.02$ fm$^{-3}$ higher in the 
QMC model.

In Table~I we give the central density, the surface energy density, 
the surface thickness and the proton radius of the droplet, as well as some 
properties of the models used, namely the incompressibility, efective nucleon 
mass and the scalar meson mass. The surface thickness, $t$, is defined as the 
width of the region where the density drops from $0.9 \rho_{B0}$ to $0.1 
\rho_{B0}$, where $\rho_{B0}$ is the baryonic density at $r=0$, after 
subtracting the background gas density. From Table~I it is seen that the 
QMC surface energy density is higher and the surface thickness
smaller. 
In fact, as a common trend in all our solutions, the surface thickness is 
$\sim 0.4-0.5$ fm lower in  the QMC model. This, 
in part, can be explained by the different values of the incompressibility 
$K$, of the models~\cite{cvinas}. Intuitively, the higher the value of $K$, 
the sharper the surface and  more similar to a theta function the density 
distribution becomes. However, as discussed in \cite{cvinas} and also 
verified in~\cite{smr}, this is not always the case when non-linear couplings 
are used. We should stress that the surface properties depend strongly not 
only on $K$ but also on $m_s$ and the effective mass $M^*$ as shown 
in~\cite{cvinas}. In fact a smaller value of $m_s$ corresponds to a larger 
range of the attractive interaction , and thus an increase of the surface 
extension.

In Fig.~4 we show the surface energy density for a droplet with total 
particle number $A=N_p+N_n=20$, with (dashed line) and without (solid line) the
electromagnetic field. For comparison we also show in this figure the
result (without the Coulomb field) for the NL1 (dotted line). As we
can see, the differences resulting from different models are much larger than 
the differences due to the inclusion of the electromagnetic field. The density
surface energy increases with the central proton fraction. In fact  for small 
proton fraction, the matter inside the droplet becomes more and more neutron 
rich and, at the same time, the density difference between the matter inside 
and outside the droplet becomes smaller. Were the matter inside and outside
the same, of course there would be no surface energy. This effect was already
observed in earlier studies of neutron star matter~\cite{baym}. Comparing the 
QMC and NL1 results, we see that the density surface energy in the 
latter is smaller,
in part due to its lower incompressibility, as discussed before.
We also see that the Coulomb field, a long range 
force, does not change much the density surface energy of the droplet, as 
expected.

In Fig.~5 we show the density surface energy for  droplets with $A=20$ (dashed 
line) and $A=40$ (solid line), both with the inclusion of the
electromagnetic field, as a function of the central proton
fraction. We see that the density surface energy increases with the total
number of particles in the droplet. This is because the systems we
studied, with $A=20$, are not big enough and, therefore, ${\cal{E}}_{surf}$ 
depends on $R$, the size of the droplet. In general ${\cal{E}}_{surf}$  is 
identified with the surface energy density associated with an infinite plane 
surface 
with a finite thickness, with the matter on one side being in the liquid phase
and on the other in the gas phase. We have checked that the density surface 
energy {\em does not} change appreciably in going from systems with $A=40$ 
to $A=60$. 

It should be stressed that when the Coulomb field is present the
solution $Y_p(r=0)=0.5$ does not exist in our formalism, since the
Coulomb field favors an increase of the neutron fraction. 

In Fig.~6 we show the proton radius as a function of the central proton 
fraction for the QMC model with (dashed line) and without (solid line) the 
contribution of the Coulomb field, and for the NL1 without the Coulomb field 
(dotted line) for a droplet with $A=20$. As the equation of state of the NL1 
model is softer than the QMC, the proton radius is larger in the 
former. It is 
interesting to notice that for $Y_p(0)$ in the interval ($0.4 - 0.45$), 
$R_p$ has a minimum. This is due to a competition between the repulsive 
Coulomb force between protons and the attractive force between protons and 
neutrons. Of course the inclusion of the Coulomb field increases the proton 
radius (due to repulsion).

The fact that for small values of the liquid proton fraction, $R_p$ is smaller
when the Coulomb field is included is because for these droplets the total 
proton fraction in the droplet is in fact smaller than the total proton 
fraction in the droplet without the Coulomb field,  as can be seen in Fig.~7 
where we plot the total proton fraction in the droplet ($Y_{tot}=N_p/A$) as
a function of the central proton fraction in the system.

Finally, in Fig.~8 we show the neutron skin thickness as a function of
the central proton fraction for a droplet with $A=20$ for the QMC
model with (dashed line) and without (solid line) the contribution of the 
Coulomb field, and for the NL1 model without the Coulomb field (dotted line). 
We 
see that the behavior is very similar in both models, the neutron skin being 
thicker in the latter because of its softer surface. We also see that the 
contribution of the Coulomb field decreases the neutron skin thickness, since 
the proton radius $R_p$ increases.    

It is worth pointing out that no results were shown for very small proton
fractions because, for the desired total number of particles (20 and 40)
we did not abtain convergence for the coupled differential equations.
In fact, for small proton fractions, when the Coulomb field is switched 
off we obtain convergence for sufficiently large droplets. As soon as we 
switch on the Coulomb field the binding energy decreases and no 
solutions appear whatever the size of the droplet. This occurs in
the QMC model, with respect to the NL1,  for smaller proton fractions.

\section {Conclusions}

In this work we have studied, within the framework of the QMC model, the 
properties of droplets which arise from the liquid-gas phase 
transition under conditions predetermined by the binodal section. The droplets 
are described using the Thomas-Fermi approximation. The results are then 
compared with the NL1.

Some of the consequences of the electromagnetic field in the droplet
formation within the QMC model, for a fixed number of particles, is to 
slightly increase the surface energy density and to decrease the central
density, increase the proton radius and decrease the neutron skin thickness
except for very small proton fractions or for almost symmetric nuclear matter.
Nevertheless, the effect of the electromagnetic field is small  in surface 
properties and the differences resulting from the use of different models are 
by far more important. The most important effect of the electromagnetic field 
is, however, to prohibit the existence of droplets with a very small 
percentage of protons, as well as very large droplets.

We have seen that the implicit density dependence of the couplings in the QMC
model has important consequences for the liquid-gas phase transition and the 
properties of the droplets. This, together with the different 
incompressibilities of the models affect sensibly the surface properties of 
the droplets. In particular, the QMC model predicts a higher density surface 
energy, a smaller proton radius and a smaller neutron skin thickness for the 
droplets than the corresponding predictions in the NL1.

We also note that while our numerical results depend on the particular model 
chosen (namely, the QMC or NL1 models), some qualitative features such as
the increase of the surface energy density, the increase of the total
proton fraction and the decrease of the neutron skin thickness with the
central proton fraction in the droplet, apply to both models.

The inclusion of finite temperature effects in droplet formation within the
QMC model is a natural extension of our calculation and work in
this direction is under consideration. We also plan to investigate the effects
of exchange terms for both the phase transition and the droplet properties. 

\acknowledgments

This work was supported in part by the University of Washington (USA), CNPq 
and FAPESP (Brazil) and FCT (Portugal) under the projects 
PRAXIS/2/2.1/FIS/451/94 and PRAXIS/P/FIS/12247/98. 
M.N. would like to thank the 
Centro de F\ii sica Te\'orica da Universidade de Coimbra for its hospitality 
and financial support during her stay in Portugal. The authors also 
acknowledge useful discussions with Dr. C.A.~Bertulani and Dr. M.B.~Pinto.


\newpage 
\begin{table}
TABLE I.  Comparison of some properties of 
 a droplet with total particle number $A=40$ and proton fraction
$N_p/A=0.5$ obtained in the Thomas-Fermi approximation with the QMC
and NL1. The values of $K$, $M^*$ and  $m_s$ are also given for both models.

\vspace{0.5cm}

\begin{center}
\begin{tabular}{lccccccc}
model &   $\rho_p(r=0)$fm$^{-3}$ & ${\cal{E}}_{surf}$ (MeV/fm$^3$)& $t$ (fm) & 
$R_p$(fm)& $K$ (MeV) & $M^*$ (MeV)& $m_s$ (MeV)\\
\hline
 QMC &0.079 & 1.15 & 2.37 & 4.00&280&754& 418\\
NL1&   0.075& 1.08& 2.79&3.94&212&535&492.25\\
\end{tabular}
\end{center}
\end{table}

\newpage

\begin{figure}
\begin{center}
\epsfig{file=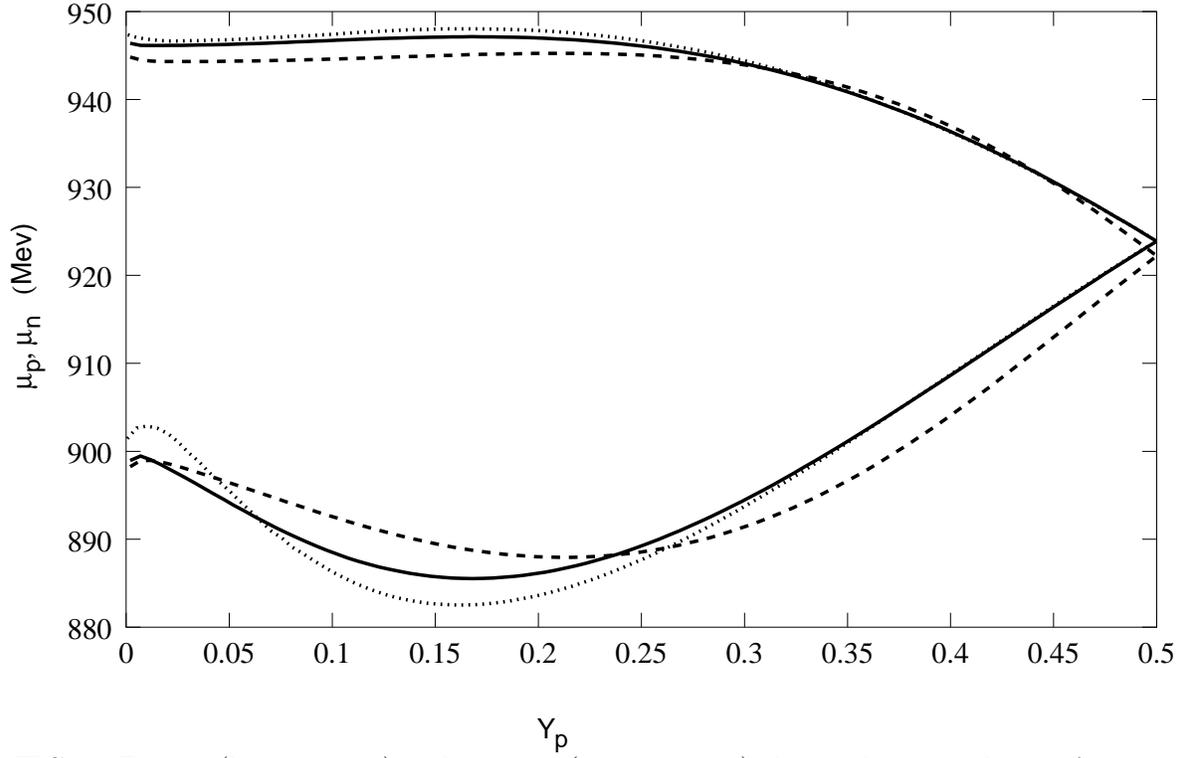,height=10cm,angle=0}
\caption{Proton (lower curves) and neutron (upper curves) chemical potentials 
as a function of the proton fraction for $P=0.1~\MeV/\fm^3$ for the QMC model 
(solid lines), for the Walecka model (dotted lines), and for the NL1 
(dashed lines).}
\label{fig1}
\end{center}
\end{figure}

\newpage
\begin{figure}
\begin{center}
\epsfig{file=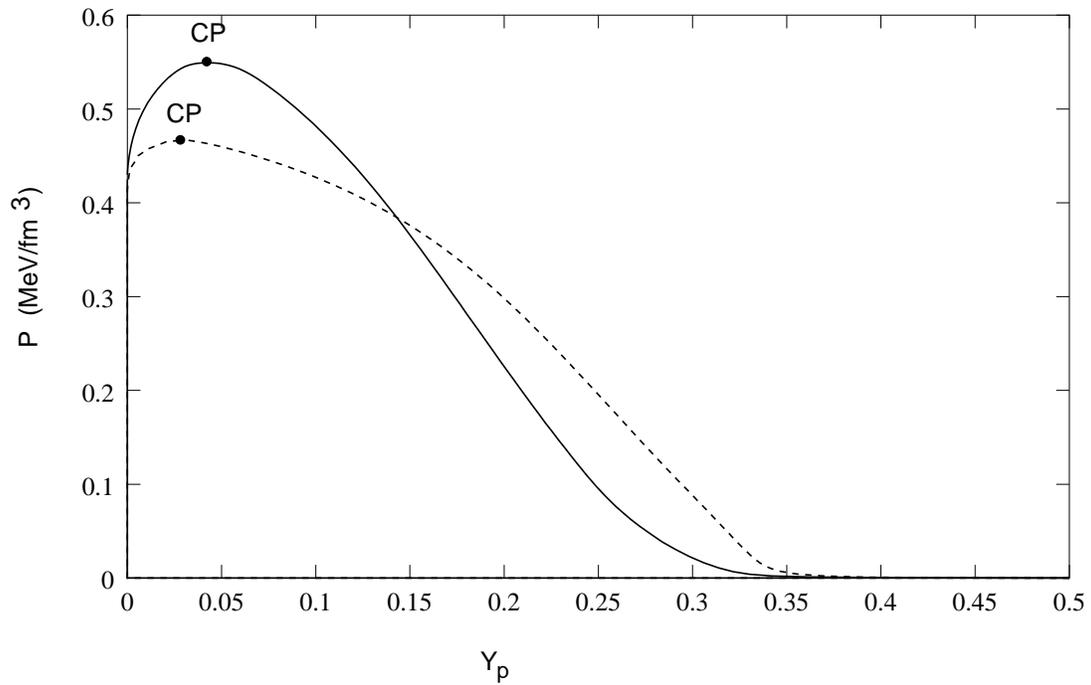,height=9cm,angle=0}
\caption{Binodal section for the QMC model (solid line) and for the 
NL1 (dashed line). The critical point (CP) is indicated; the liquid (gas) 
phase is to the right (left) of CP.}
\label{fig2}
\end{center}
\end{figure}

\newpage
\begin{figure}
\begin{center}
\epsfig{file=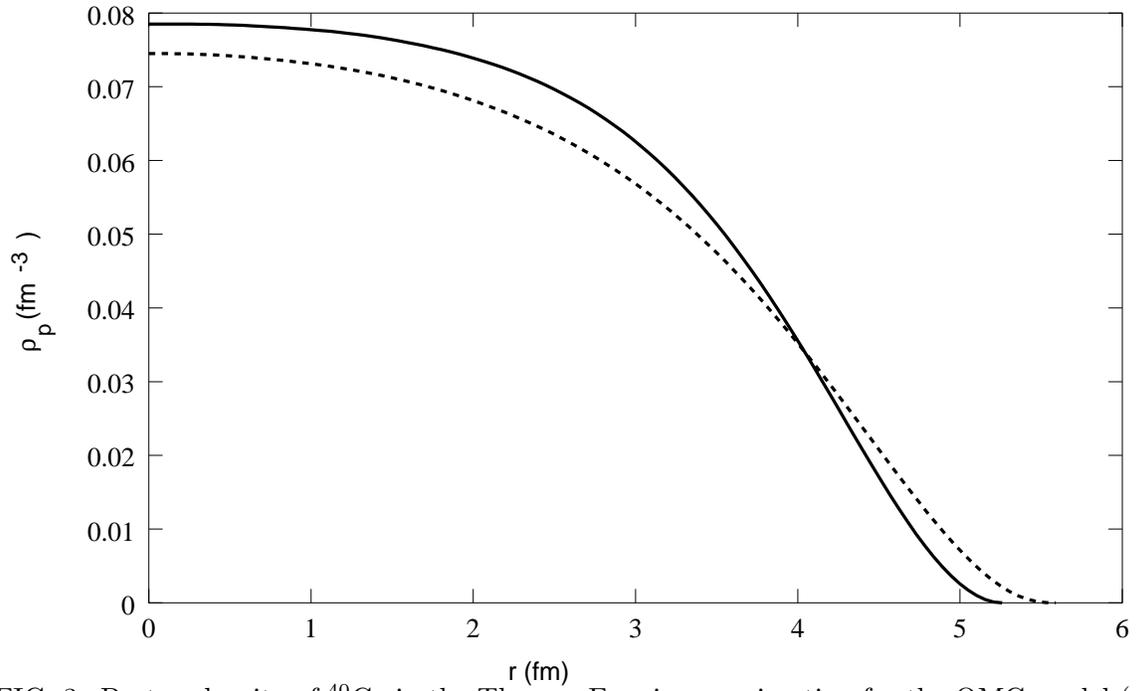,height=9cm,angle=0}
\caption{Proton density of $^{40}$Ca in the Thomas-Fermi approximation
 for the QMC model (solid line) and for the NL1 (dashed line).}
\label{fig3}
\end{center}
\end{figure}

\newpage
\begin{figure}
\begin{center}
\epsfig{file=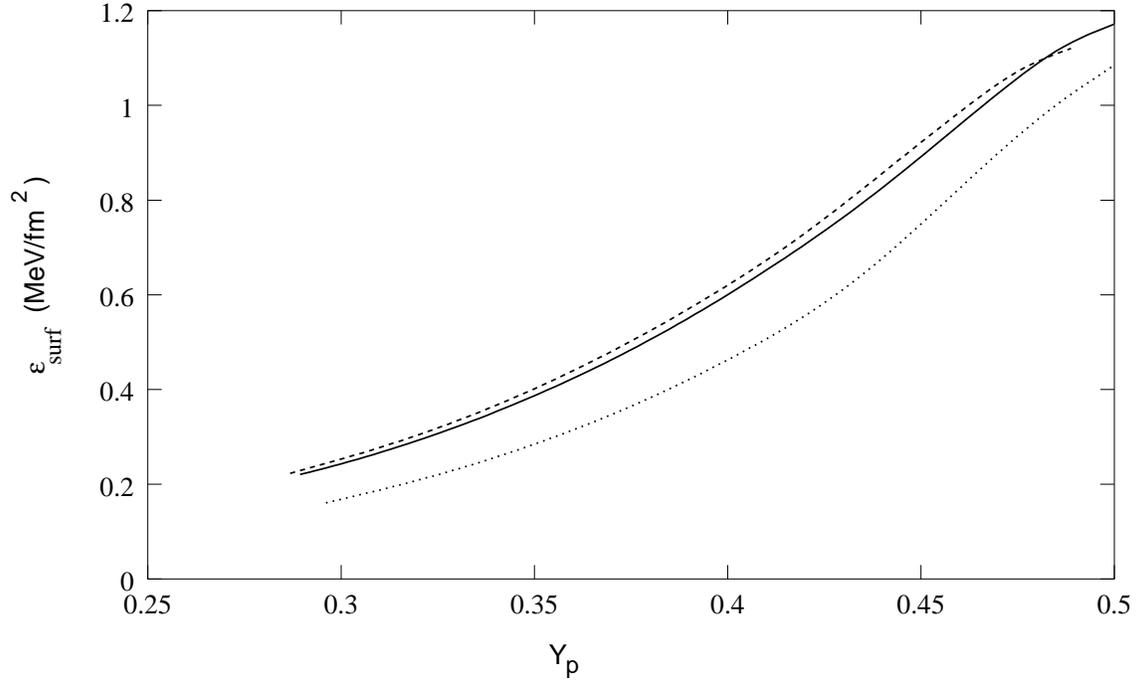,height=9cm,angle=0}
\caption{The droplet density surface energy as a function of the central
proton fraction for the QMC model with (dashed line) and without (solid line) 
the electromagnetic field, and for the NL1 without the electromagnetic field 
(dotted line). Results for a droplet with a total of 20 particles.}
\label{fig4}
\end{center}
\end{figure}

\newpage
\begin{figure}
\begin{center}
\epsfig{file=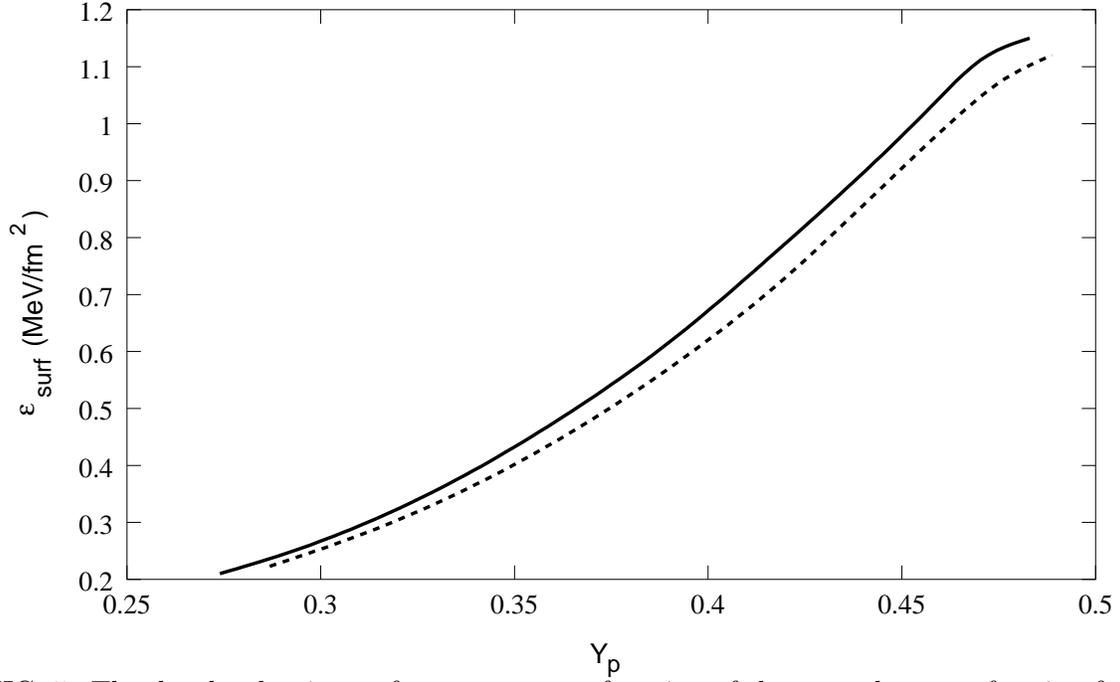,height=9cm,angle=0}
\caption{The droplet density surface energy as a function of the central
proton fraction for the QMC model with the electromagnetic field for a droplet
with a total of 40 (solid line) and 20 (dashed line) particles.}
\label{fig5}
\end{center}
\end{figure}

\newpage
\begin{figure}
\begin{center}
\epsfig{file=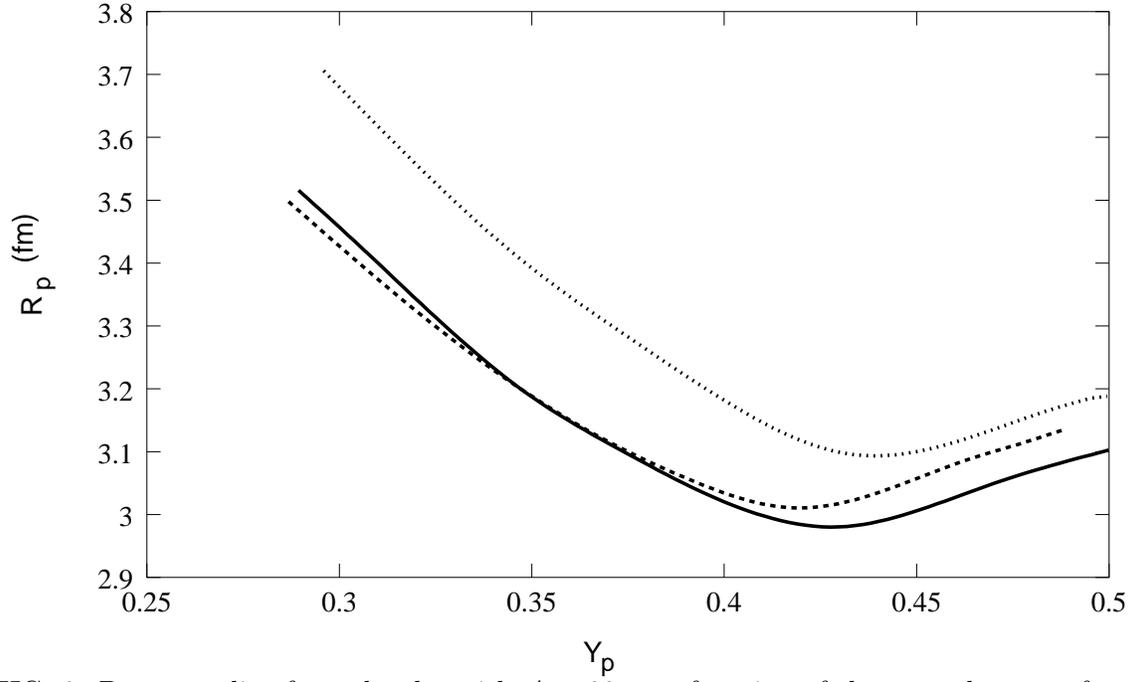,height=9cm,angle=0}
\caption{Proton radius for a droplet with $A=20$ as a function of the central 
proton fraction. The meaning of the lines is the same as in Fig. 4.}
\label{fig6}
\end{center}
\end{figure}

\newpage
\begin{figure}
\begin{center}
\epsfig{file=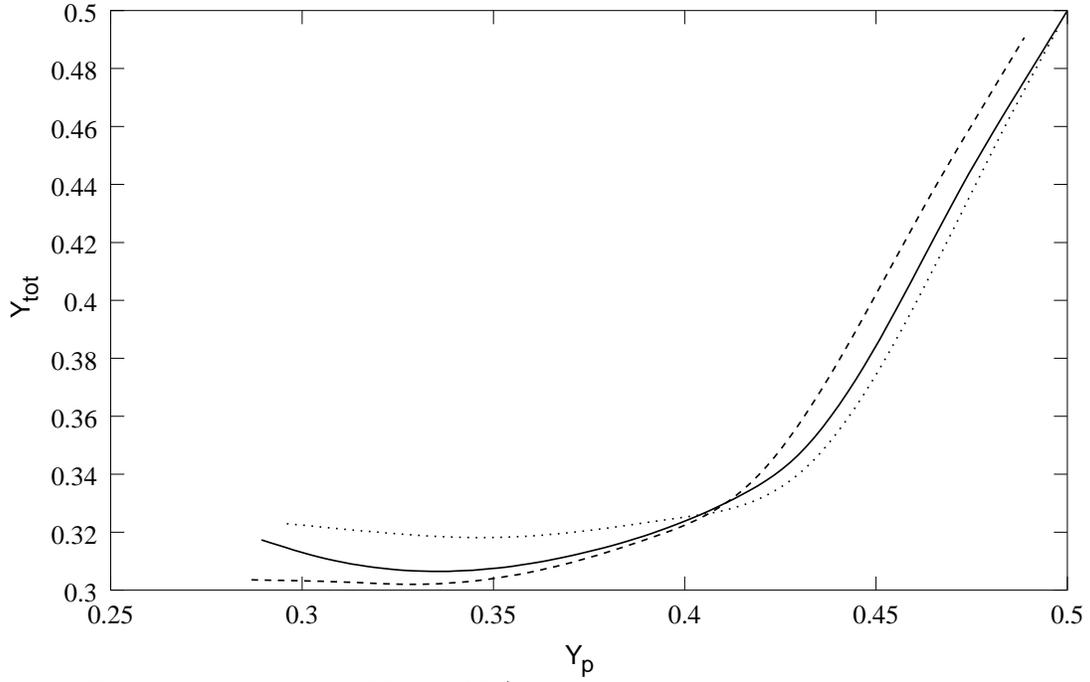,height=9cm,angle=0}
\caption{Total proton fraction $Y_{tot}=N_p/A$ in the droplet with $A=20$ as
a function of the central proton fraction. The meaning of the lines is the 
same as in Fig. 4.}
\label{fig7}
\end{center}
\end{figure}

\newpage
\begin{figure}
\begin{center}
\epsfig{file=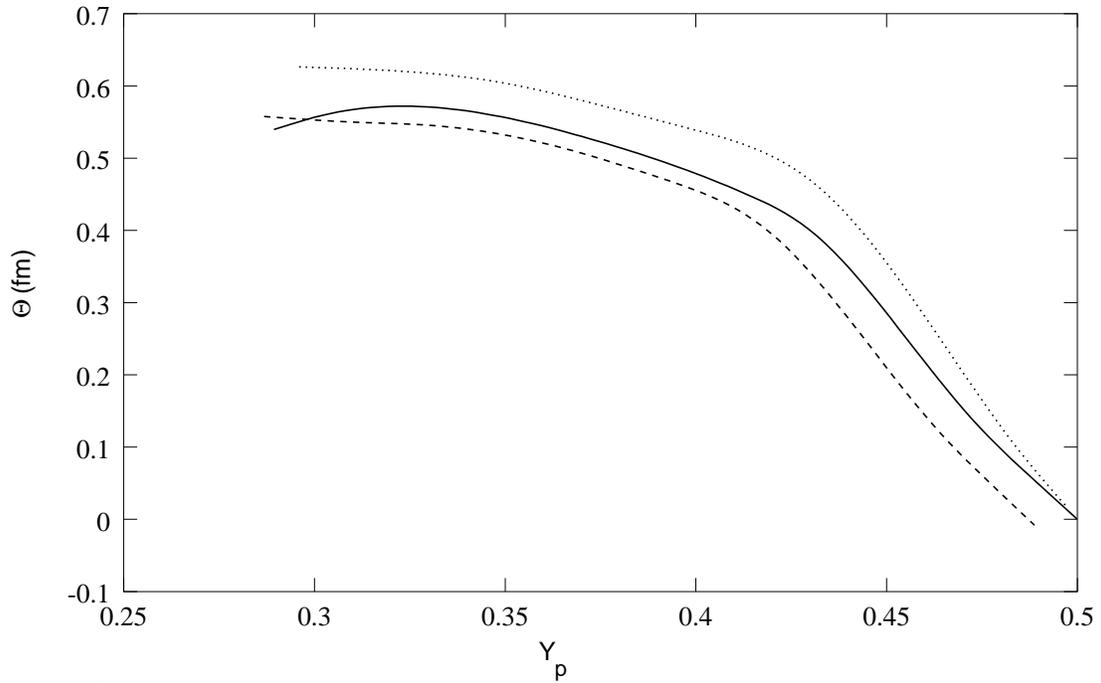,height=9cm,angle=0}
\caption{Neutron skin thickness as a function of the central proton fraction 
for a droplet with $A=20$. The meaning of the lines is the same as in Fig. 4.}
\label{fig8}
\end{center}
\end{figure}

\end{document}